\documentclass[twocolumn,pre,superscriptaddress]{revtex4}

\usepackage{dcolumn}
\usepackage{amsmath}

\usepackage{graphicx}
\usepackage{subfigure}

\newcommand{\half}{\mbox{$\frac12$}}

\newcommand{\av}[1]{\langle#1\rangle}
\newcommand{\etal}{{\it{}et~al.}}

\newcommand{\Ord}{\mathrm{O}}

\begin{document}

\title{Stochastic blockmodels and community structure in networks}
\author{Brian Karrer}
\affiliation{Department of Physics, University of Michigan, Ann Arbor, MI
48109}
\author{M. E. J. Newman}
\affiliation{Department of Physics, University of Michigan, Ann Arbor, MI
48109}
\affiliation{Center for the Study of Complex Systems, University of
Michigan, Ann Arbor, MI 48109}

\begin{abstract}
  Stochastic blockmodels have been proposed as a tool for detecting
  community structure in networks as well as for generating synthetic
  networks for use as benchmarks.  Most blockmodels, however, ignore
  variation in vertex degree, making them unsuitable for applications to
  real-world networks, which typically display broad degree distributions
  that can significantly distort the results.  Here we demonstrate how the
  generalization of blockmodels to incorporate this missing element leads
  to an improved objective function for community detection in complex
  networks.  We also propose a heuristic algorithm for community detection
  using this objective function or its non-degree-corrected counterpart and
  show that the degree-corrected version dramatically outperforms the
  uncorrected one in both real-world and synthetic networks.
\end{abstract}

\maketitle

\section{Introduction}

A stochastic blockmodel is a generative model for blocks, groups, or
communities in networks.  Stochastic blockmodels fall in the general class
of random graph models and have a long tradition of study in the social
sciences and computer
science~\cite{Holland1983,Faust1992,Anderson1992,Snijders1997,Goldenberg2009}.
In the simplest stochastic blockmodel (many more complicated variants are
possible), each of $n$ vertices is assigned to one of $K$ blocks, groups,
or communities, and undirected edges are placed independently between
vertex pairs with probabilities that are a function only of the group
memberships of the vertices.  If we denote by $g_i$ the group to which
vertex~$i$ belongs, then we can define a $K\times K$
matrix~$\boldsymbol{\psi}$ of probabilities such that the matrix element
$\psi_{g_ig_j}$ is the probability of an edge between vertices $i$ and~$j$.

While simple to describe, this model can produce a wide variety of
different network structures.  For example, a diagonal probability matrix
would produce networks with disconnected components, while the addition of
small off-diagonal elements would generate conventional ``community
structure''---a set of communities with dense internal connections and
sparse external ones.  Other choices of probability matrix can generate
core-periphery, hierarchical, or multipartite structures, among others.
This versatility, combined with analytic tractability, has made the
blockmodel a popular tool in a number of contexts.  For instance, the
planted partition model~\cite{Condon99}, which is equivalent to the model
above with a specific parametrization of the matrix~$\boldsymbol{\psi}$, is
widely used as a theoretical testbed for graph partitioning and community
detection algorithms~\cite{DDDA05,Fortunato2010}.

Another important application, and the one that is the primary focus of
this paper, is the fitting of blockmodels to empirical network data as a
way of discovering block structure, an approach referred to in the social
networks literature as \textit{a posteriori}
blockmodeling~\cite{Snijders1997}.  A number of ways of performing the
fitting have been suggested, including some that make use of techniques
from physics~\cite{Hastings2006,Hofman2008}.  \textit{A posteriori}
blockmodeling can be thought of as a method for community structure
detection in networks~\cite{Fortunato2010}, though blockmodeling is
considerably more general than traditional community detection methods,
since it can detect many forms of structure in addition to simple
communities of dense links.  Moreover, it has the desirable property (not
shared by most other approaches) of asymptotic consistency under certain
conditions~\cite{Bickel2009}, meaning that if applied to networks that were
themselves generated from the same blockmodel, the method can correctly
recover the block structure.

Unfortunately, however, the simple blockmodel described above does not work
well in many applications to real-world networks.  The model is not
flexible enough to generate networks with structure even moderately similar
to that found in most empirical network data, meaning that \textit{a
  posteriori} fits to such data often give poor results~\cite{note1}.  Just
as the fitting of a straight line to intrinsically curved data is likely to
miss important features of the data, so a fit of the simple stochastic
blockmodel to the structure of a complex network is likely to miss much
and, as we will show, can in some cases give radically incorrect answers.

Attempts to overcome these problems by extending the blockmodel have
focused particularly on the use of (more complicated) $p^*$ or exponential
random graph models, but while these are conceptually appealing, they
quickly lose the analytic tractability of the original blockmodel as their
complexity increases.  Other recent attempts to extend blockmodels take the
flavor of mixture models that allow vertices to participate in overlapping
groups~\cite{latouche2009} or to have mixed
membership~\cite{Airoldi2008,Yongjin2010}.

In this paper we adopt a different approach, considering a simple and
apparently minor extension of the classic stochastic blockmodel to include
heterogeneity in the degrees of vertices.  Despite its innocuous
appearance, this extension turns out to have substantial effects, as we
will see.  A number of previous authors have considered similar extensions
of blockmodels.  As early as 1987, Wang and Wong~\cite{Wang1987} proposed a
stochastic blockmodel for directed simple graphs incorporating arbitrary
expected in- and out-degrees, along with a selection of other features.
Unfortunately, this model is not solvable for its parameter values in
closed form which limits its usefulness for the types of calculations we
consider.  Several more recent works have also explored blockmodels with
various forms of degree heterogeneity~\cite{Dasgupta2004, Reichardt2007,
  Morup2009, CojaOghlan2009, Bader2010}, motivated largely by the recent
focus on degree distributions in the networks literature.  We note
particularly the currently unpublished work of Patterson and
Bader~\cite{Bader2010}, who apply a variational Bayes approach to a model
close, though not identical, to the one considered here.

In this paper we build upon the ideas of these authors but take a somewhat
different tack, focusing on the question of why degree heterogeneity in
blockmodels is a good idea.  To study this question, we develop a
degree-corrected blockmodel with closed-form parameter solutions, which
allows us more directly to compare traditional and degree-corrected models.
As we show, the incorporation of degree heterogeneity in the stochastic
blockmodel results in a model that in practice performs much better, giving
significantly improved fits to network data, while being only slightly more
complex than the simple model described above.  Although we here examine
only the simplest version of this idea, the approaches we explore could in
principle be incorporated into other blockmodels, such as the overlapping
or mixed membership models.

In outline, the paper is as follows.  We first review the ideas behind the
ordinary stochastic blockmodel to understand why degree heterogeneity
causes problems.  Then we introduce a degree-corrected version of the model
and demonstrate its use in \textit{a posteriori} blockmodeling to infer
group memberships in empirical network data, showing that the
degree-corrected model outperforms the original model both on actual
networks and on new synthetic benchmarks.  The benchmarks introduced, which
generalize previous benchmarks for community detection, may also be of
independent interest.

\section{Standard stochastic blockmodel}

In this section we review briefly the use of the original,
non-degree-corrected blockmodel, focusing on undirected networks since they
are the most commonly studied.

For consistency with the degree-corrected case we will allow our networks
to contain both multi-edges and self-edges, even though many real-world
networks have no such edges.  Like most random graph models for sparse
networks the incorporation of multi-edges and self-edges makes computations
easier without affecting the fundamental outcome significantly---typically
their inclusion gives rise to corrections to the results that are of
order~$1/n$ and hence vanishing as the size $n$ of the network becomes
large.  For networks with multi-edges, the previously-defined
probability~$\psi_{rs}$ of an edge between vertices in groups~$r$ and~$s$
is replaced by the expected number of such edges, and the actual number of
edges between any pair of vertices will be drawn from a Poisson
distribution with this mean.  In the limit of a large sparse graph, where
the probability of an edge and the expected number of edges become equal,
there is essentially no difference between the model described here and the
standard blockmodel.

With this in mind, the model we study is now defined as follows.  Let $G$
be an undirected multigraph on $n$ vertices, possibly including self-edges,
and let $A_{ij}$ be an element of the adjacency matrix of the multigraph.
Recall that the adjacency matrix for a multigraph is conventionally defined
such that $A_{ij}$ is equal to the number of edges between vertices $i$
and~$j$ when $i\ne j$, but the diagonal element~$A_{ii}$ is equal to
\emph{twice} the number of self-edges from $i$ to itself (and hence is
always an even number).

We let the number of edges between each pair of vertices (or between a
vertex and itself in the case of self-edges) be independently Poisson
distributed and define $\omega_{rs}$ to be the expected value of the
adjacency matrix element~$A_{ij}$ for vertices~$i$ and~$j$ lying in
groups~$r$ and~$s$ respectively.  Note that this implies that the expected
number of self-edges at a vertex in group $r$ is $\half\omega_{rr}$ because
of the factor of two in the definition of the diagonal elements of the
adjacency matrix.

Now we can write the probability~$P(G|\omega, g)$ of graph~$G$ given the
parameters and group assignments as
\begin{align}
  P(G|\omega, g) &= \prod_{i <
    j}\frac{(\omega_{g_ig_j})^{A_{ij}}}{A_{ij}!}
  \exp\bigl(-\omega_{g_ig_j}\bigr) \nonumber\\
  &\qquad{}\times \prod_i
   \frac{\bigl(\tfrac12\omega_{g_ig_i}\bigr)^{A_{ii}/2}}{(A_{ii}/2)!}
   \exp\bigl(-\tfrac12\omega_{g_ig_i}\bigr).
\label{eq:likely1}
\end{align}  

Given that $A_{ij}=A_{ji}$ and $\omega_{rs}=\omega_{sr}$,
Eq.~\eqref{eq:likely1} can after a small amount of manipulation be
rewritten in the more convenient form
\begin{align}
P(G|\omega, g) &= {1\over\prod_{i<j} A_{ij}!
  \prod_i 2^{A_{ii}/2} (A_{ii}/2)!} \nonumber\\
  &\qquad{}\times \prod_{rs} \omega_{rs}^{m_{rs}/2}
   \exp\bigl(-\tfrac12 n_r n_s\omega_{rs}\bigr),
\label{eq:likely2}
\end{align}
where $n_r$ is the number of vertices in group~$r$ and
\begin{equation}
m_{rs} = \sum_{ij} A_{ij} \delta_{g_i,r}\delta_{g_j,s},
\label{eq:defsmrs}
\end{equation}
which is the total number of edges between groups $r$ and group~$s$, or
twice that number if $r=s$.

Our goal is to maximize this probability with respect to the unknown model
parameters~$\omega_{rs}$ and the group assignments of the vertices.  In
most cases, it will in fact be simpler to maximize the logarithm of the
probability (whose maximum is in the same place).  Neglecting constants and
terms independent of the parameters and group assignments
(i.e.,~independent of $\omega_{rs}$, $n_r$, and~$m_{rs}$), the logarithm is
given by
\begin{equation}
\log P(G|\omega, g)
  = \sum_{rs} (m_{rs} \log \omega_{rs} - n_r n_s \omega_{rs}).
\label{eq:loglikely0}
\end{equation}

We will maximize this expression in two stages, first with respect to the
model parameters~$\omega_{rs}$, then with respect to the group
assignments~$g_i$.  The maximum-likelihood values $\hat{\omega}_{rs}$ of
the model parameters (where hatted variables indicate maximum-likelihood
estimates) are found by simple differentiation to be
\begin{equation}
\hat{\omega}_{rs} = {m_{rs}\over n_r n_s},
\end{equation}
and the value of Eq.~\eqref{eq:loglikely0} at this maximum is $\log
P(G|\hat{\omega}, g) = \sum_{rs} m_{rs} \log (m_{rs}/n_r n_s) - 2m$, where
$m=\frac12\sum_{rs} m_{rs}$ is the total number of edges in the network.
Dropping the final constant, we define the unnormalized log-likelihood for
the group assignment~$g$:
\begin{equation}
\mathcal{L}(G|g) = \sum_{rs} m_{rs} \log {m_{rs}\over n_rn_s}.
\label{eq:loglikely1}
\end{equation}
The maximum of this quantity with respect to the group assignments now
tells us the most likely set of assignments~\cite{CMN08,Bickel2009}.  In
effect, Eq.~\eqref{eq:loglikely1} gives us an objective or quality function
which is large for ``good'' group assignments and small for ``poor'' ones.
Many such objective functions have been defined elsewhere in the literature
on community detection and graph partitioning, but
Eq.~\eqref{eq:loglikely1} differs from most other choices in being derived
from first principles, rather than heuristically motivated or simply
proposed \textit{ad hoc}.

Equation~\eqref{eq:loglikely1} has an interesting information-theoretic
interpretation.  By adding and dividing by constant factors of the total
number of vertices and edges the equation can be written in the alternative
form
\begin{equation}
\mathcal{L}(G|g) = \sum_{rs} {m_{rs}\over2m}
                     \log {m_{rs}/2m\over n_r n_s/n^2},
\label{eq:kl}
\end{equation}
where again we have neglected irrelevant constants.  Now imagine, for a
given set of group assignments, that we choose an edge uniformly at random
from our network, and let $X$ be the group assignment at one (randomly
selected) end of the edge and $Y$ be the group assignment at the other end
of the edge.  The probability distribution of the variables~$X$ and~$Y$ is
then $p_K(X = r, Y = s) = p_K(r,s) = m_{rs}/2m$, which appears twice in the
above expression.  The remaining terms in the denominator of the logarithm
in~\eqref{eq:kl} are equal to the expected value of the same probability in
a network with the same group assignments but different edges, the edges
now being placed completely at random without regard for the groups.  Call
this second distribution~$p_1(r,s)$.  Equation~\eqref{eq:kl} can then be
written
\begin{equation}
\mathcal{L}(G|g) = \sum_{rs} p_K(r,s) \log \frac{p_K(r,s)}{p_1(r,s)},
\end{equation}  
which is the well-known Kullback--Leibler divergence between the
probability distributions $p_K$ and~$p_1$~\cite{Cover2006}.

The Kullback--Leibler divergence is not precisely a distance measure since
it's not symmetric in $p_K$ and~$p_1$.  However, if the logarithms are
taken base~2 then it measures the expected number of extra bits required to
encode $X$ and $Y$ if $p_1$ is mistakenly used as the distribution for $X$
and $Y$ instead of the assumed true distribution~$p_K$.  So intuitively it
can be considered as measuring how far $p_K$ is from~$p_1$.  The most
likely group assignments under the ordinary stochastic blockmodel are then
those assignments that require the most information to describe starting
from a model that does not have group structure.

This type of approach, in which one constructs an objective function that
measures the difference between an observed quantity and the expected value
of the same quantity under an appropriate null model, is common in work on
community detection in networks.  One widely used objective function is the
so-called modularity:
\begin{equation}
Q(g) = \frac{1}{2m} \sum_{ij}[A_{ij} - P_{ij}]\delta(g_i, g_j),
\end{equation}
where $A_{ij}$ is an element of the adjacency matrix and $P_{ij}$ is the
expected value of the same element under some null model.  The null model
assumed in our blockmodel calculation is one in which $P_{ij}$ is
constant.  Making the same choice for the modularity would lead to
\begin{equation}
Q(g) = \sum_{r=1}^{K}[p_K(r,r) - p_1(r,r)].
\end{equation}
The modularity, however, is not normally used this way and for good reason.
This null model, corresponding to a multigraph version of the
Erd\H{o}s--R\'enyi random graph, produces highly unrealistic networks, even
for networks with no community structure.  Specifically, it produces
networks with Poisson degree distributions, in stark contrast to most real
networks, which tend to have broad distributions of vertex degree.  To
avoid this problem, modularity is usually defined using a different null
model that fixes the expected degree sequence to be the same as that of the
observed network.  Within this model $P_{ij} = k_ik_j/2m$ where $k_i$ is
the degree of vertex~$i$.  Then the probability distribution over the group
assignments at the end of a randomly chosen edge becomes
\begin{equation}
p_{\textrm{degree}}(X = r, Y = s) = p_{\textrm{degree}}(r,s)
  = \frac{\kappa_r}{2m}\,\frac{\kappa_s}{2m},
\end{equation}
where
\begin{equation}
\kappa_r = \sum_s m_{rs} = \sum_i k_i \delta_{g_i,r}
\label{eq:defskappa}
\end{equation}
is the total number of ends of edges, commonly called stubs, that emerge
from vertices in group~$r$, or equivalently the sum of the degrees of the
vertices in group~$r$.  (Note that Eq.~\eqref{eq:defskappa} correctly
counts two stubs for edges that both start and end in group~$r$.)  Then the
desired group assignments are given by the maximum of
\begin{eqnarray}
Q(g) = \sum_{r=1}^{K}[p_K(r,r)-p_{\textrm{degree}}(r,r)].
\end{eqnarray}
This choice of null model is found to give significantly better results
than the original uniform model because it allows for the fact that
vertices with high degree are, all other things being equal, more likely to
be connected than those with low degree, simply because they have more
edges.  From an information-theoretic viewpoint, an edge between two
high-degree vertices is less surprising than an edge between two low-degree
vertices and we get better results if we incorporate this observation in
our model.

Returning to the stochastic blockmodel, using $p_1$ instead of
$p_{\textrm{degree}}$ in the objective function causes problems similar to
those that affect the modularity.  Fits to the model may incorrectly
suggest that structure in the network due merely to the degree sequence is
a result instead of group memberships.  We will shortly see explicit
real-world cases in which such incorrect conclusions arise.  The solution
to this problem, as with the modularity, is to define a stochastic
blockmodel that directly incorporates arbitrary heterogeneous degree
distributions.

\section{Degree-corrected stochastic blockmodel}

In the degree-corrected blockmodel, the probability distribution over
undirected multigraphs with self-edges (again denoted~$G$) depends not only
on the parameters introduced previously but also on a new set of
parameters~$\theta_i$ controlling the expected degrees of vertices~$i$.

As before, we assume there are $K$ groups, $\omega_{rs}$~is a $K\times K$
symmetric matrix of parameters controlling edges between groups $r$
and~$s$, and $g_i$ is the group assignment of vertex~$i$.  As in the
uncorrected blockmodel, let the numbers of edges each be drawn from a
Poisson distribution, but now, following~\cite{CojaOghlan2009}
and~\cite{Bader2010}, let the expected value of the adjacency matrix
element~$A_{ij}$ be $\theta_i\theta_j\omega_{g_ig_j}$.  Then graph~$G$ has
probability
\begin{align}
P(G|\theta, \omega, g) &= \prod_{i<j}
  \frac{(\theta_i\theta_j\omega_{g_ig_j})^{A_{ij}}}{A_{ij}!}
  \exp(-\theta_i\theta_j\omega_{g_ig_j}) \nonumber\\
  &\qquad{}\times \prod_i
   \frac{\bigl(\tfrac12\theta_i^2 \omega_{g_ig_i}\bigr)^{A_{ii}/2}}%
   {(A_{ii}/2)!}
   \exp\bigl(-\tfrac12 \theta_i^2 \omega_{g_ig_i}\bigr).
\end{align}
The $\theta$ parameters are arbitrary to within a multiplicative constant
which is absorbed into the $\omega$ parameters.  Their normalization can be
fixed by imposing the constraint
\begin{equation}
\sum_{i} \theta_i \delta_{g_i,r} = 1
\label{eq:sumrule}
\end{equation}
for all groups~$r$, which makes $\theta_i$ equal to the probability that an
edge connected to the community to which $i$ belongs lands on~$i$ itself.
With this constraint, the probability~$P(G|\theta, \omega, g)$ can be
simplified to the more convenient form
\begin{align}
P(G|\theta, \omega, g) 
  &= {1\over\prod_{i<j} A_{ij}!
  \prod_i 2^{A_{ii}/2} (A_{ii}/2)!} \nonumber\\
  &\qquad{}\times \prod_{i} \theta_i^{k_i}
   \prod_{rs} \omega_{rs}^{m_{rs}/2}
    \exp\bigl(-\tfrac12\omega_{rs}\bigr),
\label{eq:dclikely1}
\end{align}
with $k_i$ being the degree of vertex~$i$ as previously and $m_{rs}$
defined as in Eq.~\eqref{eq:defsmrs}.  As before, rather than maximizing
this probability, it is more convenient to maximize its logarithm, which,
ignoring constants, is
\begin{equation}
\log P(G|\theta,\omega,g)
  = 2\sum_i k_i \log\theta_i + \sum_{rs} (m_{rs} \log\omega_{rs}
    - \omega_{rs}).
\label{eq:dcloglikely1}
\end{equation}
Allowing for the constraint~\eqref{eq:sumrule}, the maximum-likelihood
values of the parameters $\theta_i$ and $\omega_{rs}$ are then given by
\begin{equation}
\hat{\theta}_i = \frac{k_i}{\kappa_{g_i}}, \qquad
\hat{\omega}_{rs} = m_{rs},
\label{eq:mlsoln}
\end{equation}
where $\kappa_r$ is the sum of the degrees in group~$r$ as before (see
Eq.~\eqref{eq:defskappa}).  This maximum-likelihood parameter estimate has
the appealing property of preserving the expected numbers of edges between
groups and the expected degree sequence of the network.  To see this, let
$\av{x}$ be the average of $x$ in the ensemble of graphs with
parameters~\eqref{eq:mlsoln}.  Then the expected number of edges between
groups $r$ and~$s$ is
\begin{equation}
\sum_{ij} \av{A_{ij}} \delta_{g_i,r}\delta_{g_j,s}
  = \sum_{ij} {k_i k_j m_{g_ig_j}\over\kappa_{g_i}\kappa_{g_j}}
     \delta_{g_i,r}\delta_{g_j,s}
  = m_{rs},
\end{equation}
where we have made use of Eq.~\eqref{eq:defskappa}.  Similarly, the average
degree of vertex~$i$ in the ensemble is
\begin{align}
\sum_j \av{A_{ij}}
  &= \sum_j \hat\theta_i\hat\theta_j\hat\omega_{g_ig_j}
   = {k_i\over\kappa_{g_i}} \sum_j {k_j\over\kappa_{g_j}} m_{g_ig_j}
     \nonumber\\
  &= {k_i\over\kappa_{g_i}} \sum_j \sum_r {k_j\over\kappa_r} m_{g_i,r}
     \delta_{g_j,r}
     \nonumber\\
  &= {k_i\over\kappa_{g_i}} \sum_r m_{g_i,r} = k_i.
\end{align}
Traditional blockmodels, by contrast, preserve only the expected value of
the matrix~$m_{rs}$ and not the expected degree---every vertex in group~$r$
in the traditional blockmodel has the same expected degree $\sum_j
m_{r,g_j}/(n_r n_{g_j}) = \kappa_r/n_r$.

Substituting Eq.~\eqref{eq:mlsoln} into Eq.~\eqref{eq:dcloglikely1}, the
maximum of $\log P(G|\theta,\omega,g)$ for the degree-corrected blockmodel
is
\begin{equation}
\log P(G|\theta,\omega,g)
  = 2\sum_i k_i \log {k_i\over\kappa_{g_i}}
     + \sum_{rs} m_{rs} \log m_{rs} - 2m.
\label{eq:logpg}
\end{equation}
where as before $m$ is the total number of edges in the network.  The first
term in this expression can be rewritten as
\begin{align}
2\sum_i k_i \log {k_i\over\kappa_{g_i}}
  &= 2\sum_i k_i \log k_i - 2\sum_i \sum_r k_i \delta_{g_i,r} \log \kappa_r
     \nonumber\\
  &\hspace{-3em} {} = 2\sum_i k_i \log k_i - \sum_r \kappa_r \log \kappa_r
     - \sum_s \kappa_s \log \kappa_s \nonumber\\
  &\hspace{-3em} {} = 2\sum_i k_i \log k_i
                      - \sum_{rs} m_{rs} \log \kappa_r\kappa_s,
\end{align}
where we have again made use of Eq.~\eqref{eq:defskappa}.  Substituting
back into Eq.~\eqref{eq:logpg} and dropping overall constants then gives us
an unnormalized log-likelihood function of
\begin{equation}
\mathcal{L}(G|g) = \sum_{rs} m_{rs} \log {m_{rs}\over\kappa_r\kappa_s}.
\label{eq:loglikely2}
\end{equation}
Notice that the only difference between this degree-corrected
log-likelihood and the uncorrected log-likelihood of
Eq.~\eqref{eq:loglikely1} is the replacement of the number~$n_r$ of
vertices in each group by the number~$\kappa_r$ of stubs.  Minor though
this replacement may seem, however, it has a big effect, as we will shortly
see.

As before, we can interpret the optimization of the objective
function~\eqref{eq:loglikely2} through the lens of information theory.
Adding and multiplying by constant factors allows us to write the
log-likelihood in the form
\begin{equation}
\mathcal{L}(G|g) = \sum_{rs} \frac{m_{rs}}{2m}
          \log \frac{m_{rs}/2m}{(\kappa_r/2m)(\kappa_s/2m)},
\end{equation} 
which is the Kullback--Leibler divergence between $p_K$ and
$p_{\textrm{degree}}$.  Alternatively, noting that $p_{\textrm{degree}}$ is
the product of the marginal distributions~$\sum_r p_K(r,s)$ and $\sum_s
p_K(r,s)$, this particular form of divergence can also be thought of as the
mutual information of the random variables representing the group labels at
either end of a randomly chosen edge.  Loosely speaking, the best fit to
the degree-corrected stochastic blockmodel gives the group assignment that
is most surprising compared to the null model with given expected degree
sequence, whereas the ordinary stochastic blockmodel gives the group
assignment that is most surprising compared to the Erd\H{o}s--R\'enyi
random graph.

Information-theoretic quantities have been proposed previously as possible
objective functions for community detection or clustering.
Dhillon~\etal~\cite{Dhillon2003}, for instance, used mutual information as
an objective function for clustering bipartite graphs, as part of an
approach they call ``information-theoretic co-clustering.''
Equation~\eqref{eq:loglikely2} is also somewhat reminiscent of an objective
function of Reichardt~\etal~\cite{Reichardt2007} which, if translated into
our terminology and adapted to undirected networks, is equivalent to the
total variation distance between $p_K$ and~$p_{\textrm{degree}}$, variation
distance being an alternative measure of the distance between two
probability distributions.  While the variation distance and the
Kullback--Leibler divergence are related, both falling in the class of
so-called $f$-divergences, the optimization of variation distance does not,
to our knowledge, correspond to maximizing the likelihood of any generative
model, and there are significant benefits to the connection with generative
models.  In particular, one can easily create networks from the ensemble of
our model and in addition the connection to generative processes means that
\textit{a posteriori} blockmodeling fits into standard frameworks for
statistical inference, which are well studied and understood in other
contexts.

Equation~\eqref{eq:loglikely2} could also be used as a measure of
assortative mixing among discrete vertex characteristics in
networks~\cite{Newman03c,Newman02f}.  In a network such as a social
network, where connections between individuals can depend on
characteristics such as nationality, race, or gender, our objective
function could be used, for instance, to quantify which of several such
characteristics is more predictive of network structure.

A useful property of the objective function in Eq.~\eqref{eq:loglikely2}
when used for \textit{a posteriori} blockmodeling is that it is possible to
quickly compute the change in the log-likelihood when a single vertex
switches groups.  When a vertex changes groups from $r$ to $s$ only
$\kappa_r$, $\kappa_s$, $m_{rt}$, and $m_{st}$ (for any~$t$) can change
(with $m_{rs}$ symmetric).  This means that many terms cancel out of the
difference of log-likelihoods and can be ignored in the computations.

\begin{widetext}
Consider moving vertex $i$ from community~$r$ to community~$s$.  Let
$k_{it}$ be the number of edges from vertex $i$ to vertices in group~$t$
excluding self-edges, and let $u_i$ be the number of self-edges for
vertex~$i$.  These quantities are the same for all possible moves of
vertex~$i$.  Define $a(x) = 2x \log x$ and $b(x) = x \log x$ where $a(0)
= 0$ and $b(0)=0$.  Then the change in the log-likelihood can be written:
\begin{align}
\Delta\mathcal{L} &= \sum_{t \neq r,s} \bigl[a(m_{rt} + k_{it}) - a(m_{rt})
  + a(m_{st} + k_{it}) - a(m_{st})\bigr] + a(m_{rs}
  + k_{ir}-k_{is})-a(m_{rs}) \nonumber\\
  &\qquad {} + b(m_{rr} - 2(k_{ir} + u_i)) - b(m_{rr}) + b(m_{ss} +
  2(k_{is}
   + u_i))-b(m_{ss}) - a(\kappa_r - k_i) + a(\kappa_r) - a(\kappa_s + k_i)
   + a(\kappa_s).
\end{align}
\end{widetext}

This quantity can be evaluated in time $\Ord(K + \left\langle k
\right\rangle)$ on average and finding the $s$ that gives the maximum
$\Delta\mathcal{L}$ for given~$i$ and~$r$ can thus be done in time
$\Ord(K(K + \left\langle k \right\rangle))$.  Because these computations
can be done quickly for a reasonable number of communities, local vertex
switching algorithms, such as single-vertex Monte Carlo, can be implemented
easily.  Monte Carlo, however, is slow, and we have found competitive
results using a local heuristic algorithm similar in spirit to the
Kernighan--Lin algorithm used in minimum-cut graph
partitioning~\cite{KL70}.

Briefly, in this algorithm we divide the network into some initial set of
$K$ communities at random.  Then we repeatedly move a vertex from one group
to another, selecting at each step the move that will most increase the
objective function---or least decrease it if no increase is
possible---subject to the restriction that each vertex may be moved only
once.  When all vertices have been moved, we inspect the states through
which the system passed from start to end of the procedure, select the one
with the highest objective score, and use this state as the starting point
for a new iteration of the same procedure.  When a complete such iteration
passes without any increase in the objective function, the algorithm ends.
As with many deterministic algorithms, we have found it helpful to run the
calculation with several different random initial conditions and take the
best result over all runs.

\section{Results}
We have tested the performance of the degree-corrected and uncorrected
blockmodels in applications both to real-world networks with known
community assignments and to a range of synthetic
(i.e.,~computer-generated) networks.  We evaluate performance by
quantitative comparison of the community assignments found by the
algorithms and the known assignments.  As a metric for comparison we use
the normalized mutual information, which is defined as
follows~\cite{DDDA05}.  Let $n_{rs}$ be the number of vertices in community
$r$ in the inferred group assignment and in community $s$ in the true
assignment.  Then define $p(X=r, Y=s) = n_{rs}/n$ to be the joint
probability that a randomly selected vertex is in $r$ in the inferred
assignment and $s$ in the true assignment.  Using this joint probability
over the random variables $X$ and $Y$, the normalized mutual information is
\begin{equation}
\textit{NMI}(X,Y) = \frac{2\,\textit{MI}(X,Y)}{H(X)+H(Y)},
\end{equation}
where $\textit{MI}(X,Y)$ is the mutual information and $H(Z)$ is the
entropy of random variable~$Z$.  The normalized mutual information measures
the similarity of the two community assignments and takes a value of one if
the assignments are identical and zero if they are uncorrelated.
A~discussion of this and other measures can be found in
Ref.~\cite{Meila2007}.

\subsection{Empirical networks}

We have tested our algorithms on real-world networks ranging in size from
tens to tens of thousands of vertices.  In networks with highly homogeneous
degree distributions we find little difference in performance between the
degree-corrected and uncorrected blockmodels, which is expected since for
networks with uniform degrees the two models have the same likelihood up to
an additive constant.  Our primary concern, therefore, is with networks
that have heterogeneous degree distributions, and we here give two examples
that show the effects of heterogeneity clearly.

The first example, widely studied in the field, is the ``karate club''
network of Zachary~\cite{Zachary1977}.  This is a social network
representing friendship patterns between the 34 members of a karate club at
a US university.  The club in question is known to have split into two
different factions as a result of an internal dispute, and the members of
each faction are known.  It has been demonstrated that the factions can be
extracted from a knowledge of the complete network by many community
detection methods.

\begin{figure}
\begin{center}
\subfigure[~Without degree correction]{%
\includegraphics[width=8cm]{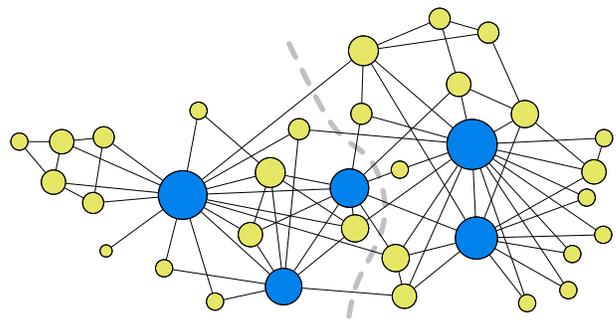}}
\subfigure[~With degree-correction]{%
\includegraphics[width=8cm]{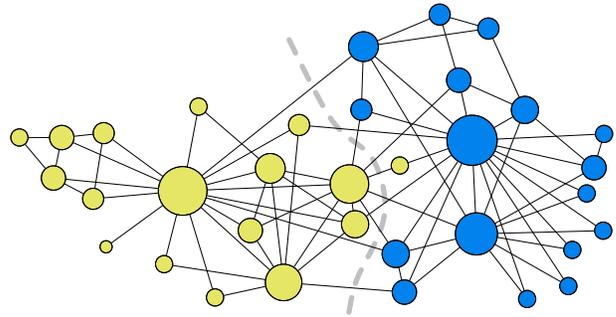}}
\end{center}
\caption{Divisions of the karate club network found using the
  (a)~uncorrected and (b)~corrected blockmodels.  The
  size of a vertex is proportional to its degree and vertex color reflects
  inferred group membership.  The dashed line indicates the split observed
  in real life.
\label{fig:karate}}
\end{figure}

Applying our inference algorithms to this network, using corrected and
uncorrected blockmodels with $K=2$, we find the results shown in
Fig.~\ref{fig:karate}.  As pointed out also by other
authors~\cite{Bickel2009,Rosvall2007}, the non-degree-corrected blockmodel
fails to split the network into the known factions (indicated by the dashed
line in the figure), instead splitting it into a group composed of
high-degree vertices and another of low.  The degree-corrected model, on
the other hand, splits the vertices according to the known communities,
except for the misidentification of one vertex on the boundary of the two
groups.  (The same vertex is also misplaced by a number of other commonly
used community detection algorithms.)

The failure of the uncorrected model in this context is precisely because
it does not take the degree sequence into account.  The \textit{a priori}
probability of an edge between two vertices varies as the product of their
degrees, a variation that can be fit by the uncorrected blockmodel if we
divide the network into high- and low-degree groups.  Given that we have
only one set of groups to assign, however, we are obliged to choose between
this fit and the true community structure.  In the present case it turns
out that the division into high and low degrees gives the higher likelihood
and so it is this division that the algorithm returns.  In the
degree-corrected blockmodel, by contrast, the variation of edge probability
with degree is already included in the functional form of the likelihood,
which frees up the block structure for fitting to the true communities.

Moreover it is apparent that this behavior is not limited to the case
$K=2$.  For $K=3$, the ordinary stochastic blockmodel will, for
sufficiently heterogeneous degrees, be biased towards splitting into three
groups by degree---high, medium, and low---and similarly for higher values
of~$K$.  It is of course possible that the true community structure itself
corresponds entirely or mainly to groups of high and low degree, but we
only want our model to find this structure if it is still statistically
surprising once we know about the degree sequence, and this is precisely
what the corrected model does.

\begin{figure}
\begin{center}
\subfigure[~Without degree-correction]{%
\includegraphics[width=8cm]{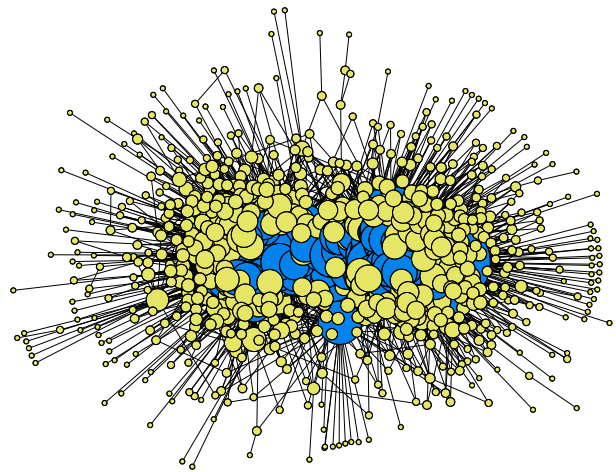}}
\subfigure[~With degree-correction]{%
\includegraphics[width=8cm]{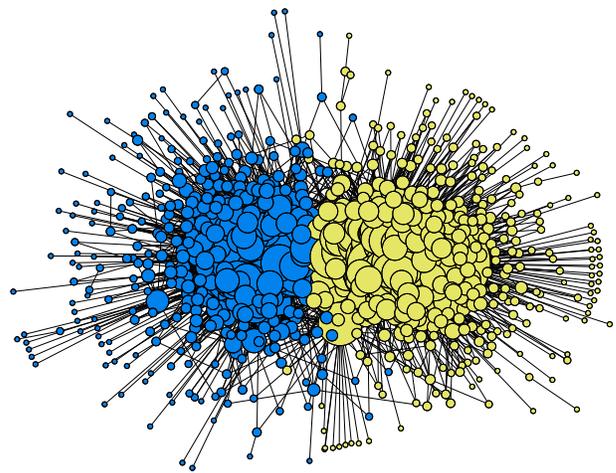}}
\end{center}
\caption{Divisions of the political blog network found using the
  (a)~uncorrected and (b)~corrected blockmodels.  The
  size of a vertex is proportional to its degree and vertex color reflects
  inferred group membership.  The division in (b) corresponds roughly to the
  division between liberal and conservative blogs given in~\cite{AG05}.
  \label{fig:blogs}}
\end{figure}

As a second real-world example we show in Fig.~\ref{fig:blogs} an
application to a network of political blogs assembled by Adamic and
Glance~\cite{AG05}.  This network is composed of blogs (i.e.,~personal or
group web diaries) about US politics and the web links between them, as
captured on a single day in 2005.  The blogs have known political leanings
and were labeled by Adamic and Glance as either liberal or conservative in
the data set.  We consider the network in undirected form and examine only
the largest connected component, which has 1222 vertices.
Figure~\ref{fig:blogs} shows that, as with the karate club, the uncorrected
stochastic blockmodel splits the vertices into high- and low-degree groups,
while the degree-corrected model finds a split more aligned with the
political division of the network.  While not matching the known labeling
exactly, the split generated by the degree-corrected model has a normalized
mutual information of $0.72$ with the labeling of Adamic and Glance,
compared with $0.0001$ for the uncorrected model.

(To make sure that these results were not due to a failure of the heuristic
optimization scheme, we also checked that the group assignments found by
the heuristic have a higher objective score than the known group
assignments, and that using the known assignments as the initial condition
for the optimization recovers the same group assignments as found with
random initial conditions.)

\subsection{Generation of synthetic networks}

We turn now to synthetic networks.  The networks we use are themselves
generated from the degree-corrected stochastic blockmodel, which is ideally
designed to play exactly this role.  (Indeed, though it is not the primary
focus of this article, we believe that the blockmodel may in general be of
use as a source of flexible and challenging benchmark networks for testing
the performance of community detection strategies.)

In order to generate networks we must first choose the values of $g$,
$\omega$, and~$\theta$.  The group assignments $g$ can be chosen in any way
we please, and we can also choose freely the values for the expected
degrees of all vertices, which fixes the $\theta$ variables according to
Eq.~\eqref{eq:mlsoln}.  Choosing the values of $\omega_{rs}$ involves a
little more work.  In principle, any set of nonnegative values is
acceptable provided it is symmetric in $r$ and~$s$ and satisfies $\sum_s
\omega_{rs} = \kappa_r$, with $\kappa_r$ as in Eq.~\eqref{eq:defskappa}.
However, because we wish to be able to vary the level of community
structure in our networks we choose $\omega_{rs}$ in the present case to
have the particular form
\begin{equation}
\omega_{rs} = \lambda \omega_{rs}^{\textrm{planted}}
  + (1-\lambda)\omega_{rs}^{\textrm{random}}.
\end{equation}
This form allows us to interpolate linearly between the values
$\omega_{rs}^{\textrm{planted}}$ and $\omega_{rs}^{\textrm{random}}$ using
the parameter~$\lambda$.  The $\omega_{rs}^{\textrm{random}}$ represents a
fully random network with no group structure; it is defined to be the
expected value of $\omega_{rs}$ in a random graph with fixed expected
degrees~\cite{CL02b}, which is simply $\omega_{rs}^{\textrm{random}} =
\kappa_r \kappa_s/2m$.

The value of $\omega_{rs}^{\textrm{planted}}$ by contrast is chosen to
create group structure.  A simple example with four groups is:
\begin{equation}
\omega^{\textrm{planted}} =
\begin{pmatrix}
  \kappa_1 & 0 & 0 & 0 \\
  0 & \kappa_2 & 0 & 0 \\
  0 & 0 & \kappa_3 & 0 \\
  0 & 0 & 0 & \kappa_4 
\end{pmatrix}.
\label{eq:diag}
\end{equation}
With this choice, all edges will be placed within communities when
$\lambda=1$ and none between communities.  When $\lambda=0$, on the other
hand, all edges will be placed randomly, conditioned on the degree
sequence, and for intermediate values of $\lambda$ we interpolate between
these two extremes in a controlled fashion.  (This model is similar to the
benchmark network ensemble previously proposed by
Lancichinetti~\cite{Lancichinetti2008}---roughly speaking it is the
``canonical ensemble'' version of the ``microcanonical'' model
in~\cite{Lancichinetti2008}.)

More complicated choices of $\omega_{rs}^{\textrm{planted}}$ are also
possible.  Examples include the core-periphery structure
\begin{equation}
\omega^{\textrm{planted}} = \begin{pmatrix}
\kappa_1-\kappa_2 & \kappa_2 \\
\kappa_2 & 0 
\end{pmatrix},
\label{eq:cp}
\end{equation}
where $\kappa_1\geq\kappa_2$.  In the case where $\kappa_1\simeq\kappa_2$
this choice also generates approximately bipartite networks, where most
edges run between the two groups and few lie inside.  Another possibility
is a hierarchical structure of the form
\begin{equation}
\omega^{\textrm{planted}} = \begin{pmatrix}
\kappa_1 - A & A & 0 \\
A & \kappa_2 - A & 0 \\
0 & 0 & \kappa_3 
\end{pmatrix},
\label{eq:hierarch}
\end{equation}
where $A \leq \textrm{min}(\kappa_1, \kappa_2)$.

In mixed models such as these, each edge in effect has a probability
$\lambda$ of being chosen from the planted structure and $1-\lambda$ of
being chosen from the null model.  Among the edges attached to a given
vertex, the expected fraction drawn from the planted structure is~$\lambda$
and the remainder are drawn randomly.

Once we have chosen our values for $g$, $\theta$, and~$\omega$, the network
generation itself is a straightforward implementation of the blockmodel: we
first draw a Poisson-distributed number of edges for each pair of groups
$r,s$ with mean~$\omega_{rs}$ (or $\frac12\omega_{rs}$ when $r=s$), then we
assign each end of an edge to a vertex in the appropriate group with
probability~$\theta_i$.

\subsection{Performance on synthetic networks}

There are two primary considerations in comparing the degree-corrected and
uncorrected blockmodels on our synthetic benchmark networks.  The first is
how close the group assignments found in our calculations are to the
planted group assignments.  The second is the performance of the heuristic
optimization algorithm.  It is possible that the maximum-likelihood group
assignment may be close to the true group assignment but that our heuristic
is unable to find it.  And if the heuristic performs better in general for
either the corrected or uncorrected blockmodel it may make comparisons
between the models unreliable: we want to claim that the degree-corrected
model gives better results than the uncorrected version because it has a
better objective function for heterogeneous networks and not because we
used a biased optimization algorithm.

To shed light on these questions we take the following approach.  For both
the degree-corrected model and the uncorrected model we perform tests with
random initial conditions and with initial conditions equal to the known
planted group structure.  The latter (planted) initializations tell us
whether the planted group assignment, or something close to it, is a local
optimum of the respective objective function---if it is, our heuristic
should find that optimum most of the time and return a final assignment
similar to the planted one.  This should be true for essentially any
reasonable heuristic, even a biased one, since the heuristic will be making
only minimal changes to the group assignments (or none at all).

For small values of~$\lambda$ we expect that the planted assignment is not
near a local maximum, but for large~$\lambda$ we would hope that it is.
Thus, if we discover in the process of running our heuristic that it is
not, it strongly suggests we have made a poor choice of objective function
(and this conclusion should hold even if the heuristic is biased).

\begin{figure*}
\begin{center}
\includegraphics[width=15cm]{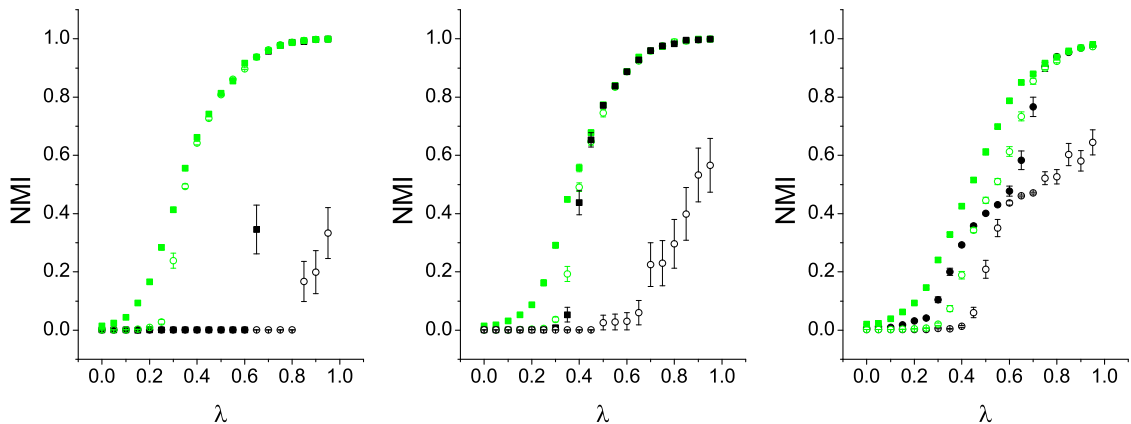}
\end{center}
\caption{The average normalized mutual information as a function of
  $\lambda$ for the three synthetic tests described in the text.  Filled
  squares and transparent circles indicate tests initialized with planted
  and random assignments respectively.  Green points denote results for the
  degree-corrected blockmodel and black for the ordinary uncorrected
  model.  The left, middle, and right panels show respectively the results
  for the two-group two-degree networks, core-periphery networks, and
  hierarchical networks.  The error bars indicate the standard error on the
  mean computed from $30$ networks per data point.}
\label{fig:examples}
\end{figure*}

The results of such tests on our synthetic networks are shown in
Fig.~\ref{fig:examples}.  We plot the normalized mutual information as a
function of $\lambda$ for various choices of planted structure.  Each data
point represents an average over $30$ networks of size $n=1000$ for both
the degree-corrected and uncorrected objective functions.  In the case of
random initializations, ten initializations were performed for each
network and we take the best result among the ten.

The left panel in the figure shows results for networks with two
communities and just two possible values of the expected degree, 10 and~30.
Each of the 1000 vertices was assigned to one of the four possible
combinations of degree and community with equal probability, and the
planted structure was chosen diagonal, as in Eq.~\eqref{eq:diag}.

The green points in the figure indicate the performance of the
degree-corrected blockmodel, while the black points are for the uncorrected
model.  Solid squares and open circles show performance starting from the
planted community structure and random assignments respectively.  Bearing
in mind that $\lambda=0$ corresponds to zero planted structure (in which
case neither algorithm should find any significant result) and that a
normalized mutual information approaching~1 indicates successful detection
of the planted structure, we can see from the figure that the
degree-corrected blockmodel significantly out-performs the uncorrected one
in this simple test.  As $\lambda$ increases from zero, the mutual
information for all algorithms rises, but the corrected model starts to
detect some signatures of the planted structure almost immediately and for
$\lambda=\frac12$ returns a normalized mutual information above 0.7 for
both initial conditions.  The uncorrected model, by contrast, finds no
planted structure at $\lambda=\frac12$ for either
initialization---including when the algorithm is initialized to the known
correct answer.  The reason for this poor performance is precisely because
of the variation in degrees: for values of $\lambda$ up to around 0.6 the
uncorrected model fits these networks better if vertices are assigned to
groups according to their degree than if they are assigned according to the
planted structure, and hence the best-fit group structure has no
correlation with the planted structure.

We have also tested our blockmodels against synthetic networks with two
other types of structure, one the core-periphery or approximately bipartite
structure of Eq.~\eqref{eq:cp} and the other the hierarchical structure of
Eq.~\eqref{eq:hierarch}.  In these examples we use a more realistic degree
distribution that approximately follows a power law with a minimum expected
degree of~10 and an exponent of~$-2.5$.  For the core-periphery networks we
randomly assign vertices to one of the two groups, while for the
hierarchical networks we fix 500 vertices to be in the first of the three
groups, put the rest randomly in the other two, and set $A =
\frac14\textrm{min}(\kappa_1,\kappa_2)$.  (It has been suggested that
choosing non-equal sizes for groups in this way presents a harder challenge
for structure detection algorithms~\cite{Danon2006a, Rosvall2007}.)

The performance of our blockmodels on these two classes of networks is
shown in the middle (core-periphery) and right (hierarchical) panels of
Fig.~\ref{fig:examples}.  Again we see that the normalized mutual
information increases with increasing $\lambda$ for all algorithms but that
the degree-corrected blockmodel performs significantly better than the
uncorrected model.  The degree-corrected model with planted assignments
consistently does the best among the four options as we would expect, and
the degree-corrected model with random initializations performs respectably
in all cases, although it's entirely possible that better performance could
be obtained with a better optimization strategy.  The performance of the
uncorrected model with random initializations, on the other hand, is quite
poor~\cite{note2}.  But perhaps the most telling comparison is the one
between the degree-corrected model with random initial assignments and the
uncorrected model with the planted assignment.  This comparison tilts the
playing field heavily in favor of the uncorrected model and yet, as
Fig.~\ref{fig:examples} shows, the degree-corrected model still performs
about as well as, and in some cases better than, the uncorrected model.

\section{Conclusions}

In this paper, we have studied how one can incorporate heterogeneous vertex
degrees into stochastic blockmodels in a simple way, improving the
performance of the models for statistical inference of group structure.
The resulting degree-corrected blockmodels can also be used as generative
models for creating benchmark networks, retaining the generality and
tractability of other blockmodels while producing degree sequences closer
to those of real networks.

We have found the performance of the degree-corrected model for inference
of group structure to be quantitatively better on both synthetic and
real-world test networks than the uncorrected model.  In networks with
substantial degree heterogeneity, the uncorrected model prefers to split
networks into groups of high and low degree, and this preference can
prevent it from finding the true group memberships.  The degree-corrected
model correctly ignores divisions based solely on degree and hence is more
sensitive to underlying structure.

It seems likely that other more sophisticated blockmodels, such as the
recently proposed overlapping and mixed membership models, would benefit
from incorporating degree sequences also.  In applications to on-line social
network data, for example, where overlapping groups are common, there is
frequently substantial degree heterogeneity and hence potentially
significant benefits to using a degree-corrected model.

The degree-corrected blockmodel is not without its faults.  For instance,
the model as described can produce an unrealistic number of zero-degree
vertices, and is also unable to model some degree sequences, such as those
in which certain values of the degree are entirely forbidden.  As a model
of real-world networks, it may also fail to accurately represent
higher-order network structure such as overrepresented network motifs or
degree correlations.  From a statistical point of view, it is also somewhat
unsatisfactory that the number of parameters in the model scales with the
size of the network, which for example prevents fits to a network of one
size being used to generate synthetic networks of another size.

But perhaps the chief current drawback of the model is that the number~$K$
of blocks or groups in the network is assumed given.  In most structure
detection problems the number of groups is not known and a complete
calculation will therefore require not only the algorithms described in
this paper but also a method for estimating~$K$.  Some previously suggested
approaches to this problem include cross-validation~\cite{Airoldi2008},
minimum description length methods using two-part or universal
codes~\cite{Rosvall2007}, maximization of a marginal
likelihood~\cite{Hofman2008}, and nonparametric Bayesian methods.  The
marginal likelihood for our degree-corrected blockmodel can be computed
explicitly if one assumes conjugate priors on the parameters---Dirichlet
for $\theta$ and gamma for~$\omega$---but then one must also choose the
parameters of those priors, called hyper\-parameters in the statistical
literature.  In principle one wants to choose values of the hyperparameters
that provide asymptotic consistency---the blockmodel should return the
correct number of groups when applied to a network generated from the same
blockmodel, at least in certain limits.  At present, however, it is not
known how to make this choice.  An alternative possibility is to note that
the blockmodel used here is equivalent to a model that generates an
ensemble of matrices with integer entries, implying potential connections
to the large statistical literature on contingency table analysis that
could be helpful in determining the number of groups in a principled
fashion.  We leave these questions for future work.

\begin{acknowledgments}
  The authors would like to thank Joel Bader and Cris Moore for useful
  conversations, and Joel Bader for sharing an early draft of
  Ref.~\cite{Bader2010}.  This work was funded in part by the National
  Science Foundation under grant DMS--0804778 and by the James S.
  McDonnell Foundation.
\end{acknowledgments}


\begin{thebibliography}{30}
\bibitem{note1} Technically this is only true if the number~$K$ of groups
  is moderate.  If $K$ can become arbitrarily large then it is possible to
  represent any network trivially as a stochastic blockmodel by making each
  vertex its own group and setting the independent probability of each edge
  to one or zero depending on whether an edge exists between the
  corresponding vertices in the empirical data.
\bibitem{note2} This algorithm's performance may not be quite as bad as the
  figure seems to indicate because for high $\lambda$ the mutual
  information values are roughly bimodal, with most values being close to
  either 0 or~1.  The mean shown is thus roughly the percentage of times
  that the uncorrected model successfully finds the planted structure.  The
  same does not apply for the other curves, where the error represents
  relatively minor variation about the mean.
\end{thebibliography}
\end{document}